\documentclass[10pt]{article}
\usepackage{graphics}
\usepackage{graphicx}
\usepackage[below]{placeins}
\usepackage{psfig}
\usepackage{epsf}

\newcommand{\beq}{\begin{equation}}
\newcommand{\eeq}{\end{equation}}
\newcommand{\beqa}{\begin{eqnarray}}
\newcommand{\eeqa}{\end{eqnarray}}
\newcommand{\forget}[1]{}

\newcommand{\SB}[1]{}

\begin{document}
\newcommand{\tabletyp}{
   \begin{table}[htb]
      \begin{center}
         \begin{tabular}{|l|c|c|c|c|c|c|} \hline
            & $P(\sigma_\mathbf{1})$ & $P(\sigma_\mathbf{0})$ &
            $P(\sigma_\mathbf{r})$ & Max $P_{e}$ & Min $P_{e}$ &
            Median $P_{e}$ \\ \hline
            Pro B-cells & 0.405 & 0.460 & 0.135 & 0.035 & 0.0002 &
            0.020 \\ \hline
            Pre B-cells & 0.395 & 0.450 & 0.155 & 0.047 & 0.003 &
            0.028 \\ \hline
            Full B-cells & 0.343 & 0.471 & 0.186 & 0.073 & 0.0007 &
            0.022 \\ \hline
        \end{tabular}
       \end{center}
       \caption{Summary of the estimated parameter values for the
       B-cell data. $P_{e}$ refers to the set of error probabilities,
       i.e., $[P^i_{1 \rightarrow 0}, P^i_{0 \rightarrow 1}]^4_{i=1}$.}
       \label{params}
   \end{table}
}

\newcommand{\tablettest}{
\begin{table}[h]
\begin{center}
  \begin{tabular}{|l|c|c|c|c|c|c|c|c|} \cline{3-8}
\multicolumn{2}{c}{ } &  \multicolumn{3}{|c|}{Discrete} &
 \multicolumn{3}{|c|}{Continuous} & \multicolumn{1}{c}{ }  \\ \hline
 Gene & Id & \ I \  & II & III & \ I \ & II & III & Literature \\ \hline
 \bf{Target profile} & - & \bf{S} & \bf{U} &\bf{U} 
 & \bf{S} & \bf{U} &  \bf{U} & \\ \cline{1-8}  
 CD20 & 99446\_at & S & U & U & S & U & U & B-Cell \\ \cline{1-8}
 Spi  & 93657\_at  & S & U & U & S & S & S & specific \\ \hline 
 \bf{Target profile} & - & \bf{U} & \bf{D} &\bf{S} 
 & \bf{U} & \bf{D} &  \bf{S} & \\ \cline{1-8}  
 Sox-4 & 160109\_at & U & D & S & U & D & S  & \\ \cline{1-8}
 lef-1 & 103628\_at & U & D & S & U & S & S  & \\ \cline{1-8}
 rag-1 & 93683\_at  & U & S & S & S & S & U  & \\ \cline{1-8}
 VpreB & 92972\_at & U & D & S & U & D & S & Pre-B \\ \cline{1-8}
 Lambda-5 & 99429\_at & U & D & S & U & D & S & specific \\ \cline{1-8}
 Il-7 receptor & 99030\_at & U & D & S & U & S & S &  \\ \cline{1-8}
 TdT & 99030\_at & U & D & S & U & S & S &  \\ \hline
 \bf{Target profile} & - & \bf{U} & \bf{S} &\bf{U} 

 & \bf{U} & \bf{S} &  \bf{U} & \\ \cline{1-8}  
 Bob-1 & 93915\_at & U & S & U & U & S & U & \\ \cline{1-8}
 CD 19 & 99945\_at & U & S & U & U & D & U & \\ \cline{1-8}
 Blnk  & 100771\_at & U & S & U & U & S & U & \\ \cline{1-8}
 Pax-5 & 96993\_at & U & S & U & U & S & U & B-lineage \\ \cline{1-8}
 Blk & 92359\_at & U & S & U & U & S & U & Specific \\ \cline{1-8}
 Mb-1 & 102778\_at & U & D & S & U & D & S & \\ \cline{1-8}
 B29 & 161012\_at & S & S & S & U & S & U & \\ \cline{1-8}
 CD24 & 100600\_at & U & S & U & U & S & U & \\ \hline
\bf{Target profile} & - & \bf{D} & \bf{S} &\bf{D} 
 & \bf{D} & \bf{S} &  \bf{D} & \\ \cline{1-8}  
 Id-1 & 100050\_at & D & S & D & D & S & D &  \\ \cline{1-8}
 Fag-1 & 97974\_at & S & D & D & D & S & D & Not in \\ \cline{1-8}
 Il-3 receptor & 94747\_at & D & S & D & D & S & D & B-lineage \\ \cline{1-8}
 CD 63 & 160493\_at & D & S & D & D & S & D & \\ \cline{1-8}
 Gata-2 & 102789\_at & D & S & D & D & D & D & \\ \hline
\end{tabular}
\end{center}
\caption{The three groups I, II and III indicate the expressional
 changes between {\em Pro-B to Pre-B}, {\em Pre-B to Mature-B}, and
{\em Pro-B to Mature-B} respectively. U stands for accepting the
hypothesis up, D for down, and S (stable) if no hypothesis could be
accepted on the 95\% confidence level.}
\label{Hugo}
\end{table}


}

\newcommand{\showhere}[1]{}    

\newcommand{\fignoiseline}{
  \begin{figure}[!htb]
  \centerline{\psfig{file=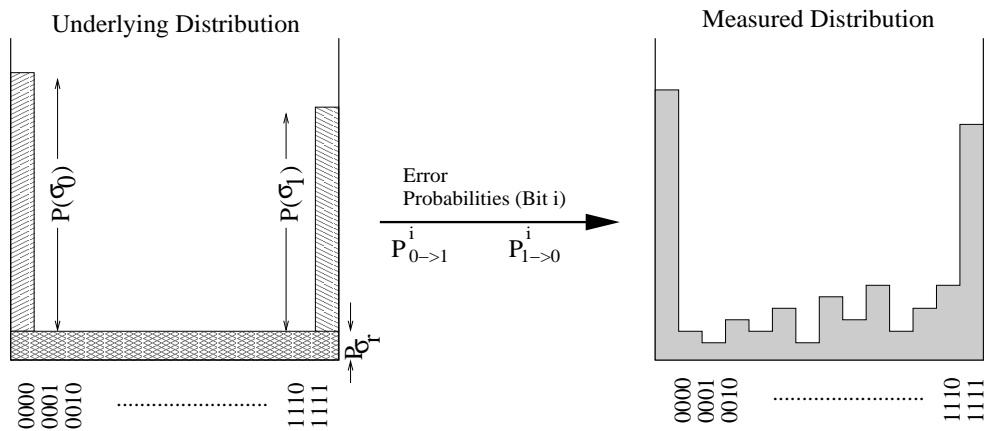,width=13cm}}
  \caption{ Schematic diagram illustrating the transition from
      underlying to observed distributions of states, in the case of
      $m=4$ samples. The underlying distribution on the left hand side
      can be described by the probabilities for each underlying state,
      $P(\sigma_\mathbf{1})$, $P(\sigma_\mathbf{0})$, and
      $P(\sigma_\mathbf{r})$ (see text). This distribution is then
      distorted by sample specific errors, $P^i_{0 \rightarrow 1}$ and
      $P^i_{1 \rightarrow 0}$, resulting in an experimentally
      observed distribution, depicted  on the right hand side.}
      \label{noiseline}
  \end{figure}
}

\newcommand{\figcorrbio}{
   \begin{figure}[!htb]
  \centerline{\psfig{file=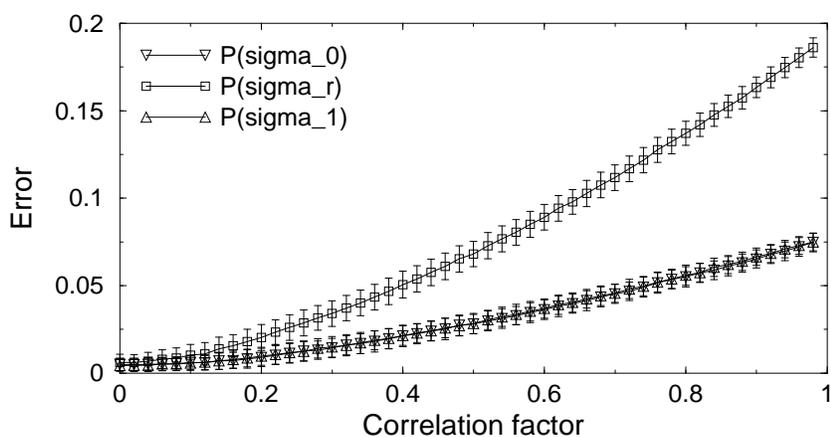,angle=270,width=11cm}}
   \caption{The average error in the estimation of the parameters
      $P(\sigma_{\mathbf{1}})$, $P(\sigma_{\mathbf{0}})$,
      $P(\sigma_{\mathbf{r}})$ are given as a function of correlation factor
      between the third and fourth bit. For correlation factors above 0.2
      the error in $P(\sigma_{\mathbf{r}})$ rises considerably.  }
   \label{globals}
   \end{figure}
}

\newcommand{\figcorrtran}{
   \begin{figure}[!htb]
   \centerline{\psfig{file=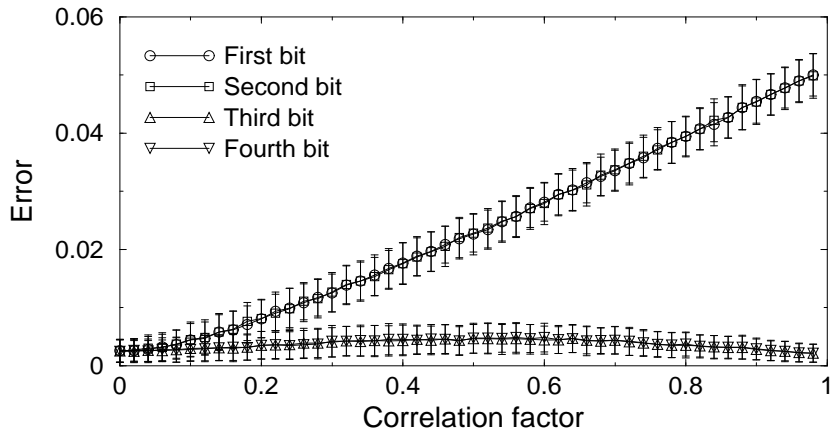,angle=270,width=11cm}}
   \caption{The average error in the estimation of the error
     probabilities $\{P^i_{0 \rightarrow 1}\}_{i=1}^{4}$. For correlation
     factors above 0.2 $P^1_{1 \rightarrow 0}$ and $P^2_{1 \rightarrow 0}$ 
     are notably raised. Patterns were these bits deviate from the
     other two are then not considered as random but rather caused by
     an error. This effect could only be avoided by introducing
     extra parameters for correlation between bits.}
   \label{flips}
   \end{figure}
}

\newcommand{\figinfoloss}{
   \begin{figure}[!htb]
   \begin{center}
   \rotatebox{-90}{\resizebox{!}{11cm}{\includegraphics{./fig2.eps}}}
   \end{center}
   \caption{Distribution of p-values in $t$-tests of different expression
      between any two varieties over all genes. The upper curve
      labeled ``Original ordering'' shows the density for the original
      assignment of samples to the three varieties. The lower curve
      shows the average of p-values for data where the assignment of
      samples to varieties is random. }
   \label{infoloss}
   \end{figure}
}

\newcommand{\appendfigsandtab} {
  \fignoiseline \vfill

  \figinfoloss \vfill

  \figcorrbio \vfill

  \figcorrtran \vfill
 
  \tabletyp \vfill 

  \tablettest
}



\begin{titlepage}

\begin{flushright}
LU TP 02-14\\
May 9, 2003
\end{flushright}

\vspace{.25in}

\LARGE

\begin{center}
  {\bf Probabilistic estimation of microarray \\ data reliability and underlying
    gene expression}\\
\vspace{.3in}
\large

{\sl S. Bilke$^*$\footnote[1]{sven@thep.lu.se}, 
     T. Breslin$^*$\footnote[2]{thomas@thep.lu.se}
 and M. Sigvardsson$^\dagger$\footnote[3]{Mikael.Sigvardsson@stemcell.lu.se}
\vspace{0.6cm}
}

$^*$Complex Systems Division, Department of Theoretical Physics\\ 
University of Lund,  S\"{o}lvegatan 14A,  SE-223 62 Lund, Sweden \\
{\tt http://www.thep.lu.se/complex/}
\vspace{0.3cm}

$^\dagger$The Laboratory for Cell Differentiation Studies\\
Department for Stem Cell Biology, BMC B12, SE-22185 Lund, Sweden \\
{\tt http://www.stemcell.lu.se/stemcell.html}
\end{center}
\vspace{0.25in}

\begin{center}
Submitted to {\it BMC Bioinformatics }
\end{center}

\large

\vspace{0.8in}

\end{titlepage}


\newpage
\section{Abstract}

{\bf Background:} The availability of high throughput methods for
measurement of mRNA concentrations makes the reliability of
conclusions drawn from the data and global quality control of samples
and hybridization important issues. We address these issues by an
information theoretic approach, applied to discretized expression
values in replicated gene expression data.

{\bf Results:} Our approach yields a quantitative measure of two
important parameter classes: First, the probability $P(\sigma | S)$
that a gene is in the biological state $\sigma$ in a certain variety,
given its observed expression $S$ in the samples of that
variety. Second, sample specific error probabilities which serve as
consistency indicators of the measured samples of each variety. The
method and its limitations are tested on gene expression data for
developing murine B-cells and a $t$-test is used as reference. On a
set of known genes it performs better than the $t$-test despite the
crude discretization into only two expression levels. The consistency
indicators, i.e. the error probabilities, correlate well with
variations in the biological material and thus prove efficient.

{\bf Conclusions:} The proposed method is effective in determining
differential gene expression and sample reliability in replicated
microarray data. Already at two discrete expression levels in each
sample, it gives a good explanation of the data and is comparable to
standard techniques.

\twocolumn
\section{Background}

A broad variety of algorithms has been developed and used to extract
biologically relevant information from gene expression data.  Among
others commonly used are visual inspection \cite{visual}, hierarchical
and k-means clustering \cite{esbb98}, self organizing maps \cite{som1,
som2} and singular value decomposition \cite{hmmcbf00, abb00}. These
methods aim  mainly at identifying predominant patterns and thus groups
of ``cooperating'' genes based on the assumption that related genes
have similar expression patterns.

Compared to the amount of work devoted to efficient methods to extract
information from the data, somewhat less attention has been paid to
the question of the reliability of the generated results. The ANOVA
analysis \cite{anova} allows estimation, and thus elimination, of some
systematic error sources.  Bootstrapping cluster analysis estimates
the stability of cluster assignments \cite{bootstrap} based on
artificial data-sets generated with ANOVA coefficients. Some authors
also considered the question of how well a certain oligo
\cite{oligo-quality}
is suited to measure the mRNA expression level of the related gene.

Some work has gone towards the ambitious task of learning topological
properties or qualitative features of the genetic regulatory network
from expression profiles, see e.g.  \cite{bayes}.  A major limiting
factor in these attempts is the comparative sparseness of available
data. It is therefore reasonable to consider reduced models, for
example a Boolean representation of the gene activity. It is known
that many biological properties, for instance stability and
hysteresis, can be modeled by the dynamics of such reduced models
\cite{kauffman1, kauffman2, huang}.

In this work we investigate the possibility of reducing complexity of
gene expression data by discretizing the expression levels. The
approach we present enables a new way of extracting biologically
relevant information from the data in the following way: A biological
variety, i.e. a biological system defined by the investigator, is
represented by several samples which are subjected to gene expression
analysis. If gene expression levels are discretized into $n$ values,
and the variety is represented by $m$ samples, the number of
observable expression states for a gene are limited to $n^m$. These
observed states $S$ are modeled as being derived from a smaller number
of underlying, biological states $\sigma$, through a measurement
process. Rather than making static assignments $S \rightarrow \sigma $
we calculate conditional probabilities $P( \sigma | S)$. The number of
possible expression profiles for a gene over a set of varieties is
limited and the probability of each expression profile is easily
calculated. Since the model we use considers both the underlying
biology and the measurement process it also generates a measure of
sample coherence in each biological variety.

We demonstrate the feasibility of this approach for a {\em binary}
discretization of gene expression. For the discretization step we use
the absent/present classification provided by the Affymetrix software
\cite{affysoft}. The outcome of our method on a data set covering gene
expression in developing murine B-cells is compared to the results of
a standard analysis. We show that even with the crude discretization
into only two expression levels the method is competitive to
statistical methods based on continuous expression levels.

\section{Methods} \label{tmodel}
\subsection{The Model}

A major step in the analysis of gene expression data is to separate
the biological content of the data from measurement and sample
specific errors. In other words given an observation, i.e. the
expression values of a gene in several samples representing the same
biological variety\footnote{In the application on which we demonstrate
the method we consider three different varieties: pro, pre, and mature
B-cells. The samples in each variety are different cell lines arrested
at the corresponding stage of development}, one wants to conclude on
the biological state $\sigma $, which generated the observation. This
can be expressed as a conditional probability \beq P(\sigma| S),
\label{target}
\eeq that a gene is in a certain biological state $\sigma$ given the
corresponding observed state $S$.

In this work we take an information theoretic point of view to
estimate this probability: The information of interest, the state
$\sigma$, is ``transmitted'' in a noisy measurement process and
potentially distorted (Figure~\ref{noiseline}).  Using Bayes' theorem,
the desired conditional probability Eq.\ (\ref{target}) can be expressed
as: \beq P(\sigma | S) = \frac{P(S | \sigma) P_{\sigma}} {P_S}.
\label{maineq}
\eeq
On the right hand side of this equation, $P(S | \sigma)$ is the
probability to observe state $S$ if the underlying biological state is
$\sigma$. In a sense, $P(S | \sigma)$  describes the noise characteristic
of the measurement process. In the following we will show how this
conditional probability, and the other probabilities on the r.h.s. of
Eq.\ (\ref{maineq}) can be estimated.

Given a set of $m$ samples representing the same biological variety,
differences in the expression level of a gene between the samples can
arise from two independent sources:

\begin{enumerate}
\item \emph{Random variation} within the variety. This may be caused
by temporal differences in response to the stimuli, slightly different
environmental conditions, genotypic differences between samples,
etc.
\label{bierr}
\item \emph{Sample specific errors}. These  are mainly caused by the
measurement process, e.g. differences in the treatment of the mRNA,
scratched arrays, and so on. However, outlier samples, cultured under
considerably different conditions, also contribute to sample specific
errors.
\label{trerr}

\end{enumerate}
A separation of these two contributions is possible only with an
appropriate model for the variation of gene expression between the
samples. In the choice of model, one has  considerable freedom
within the bounds set by biological plausibility. A limiting factor on
the biological model comes from the type and amount of available
data. The data used in this work contains only four samples for each
variety. For the model we propose this is the {\em minimum} number of
samples required to estimate the model parameters.

In the discretization of gene expression levels, we use only two
discrete values, $0$ and $1$, for the expression of a gene in a
sample. This means that the number of observable states, $S$, in a
variety consisting of $m$ samples is $2^m$. With no measurement errors
we could immediately conclude on the underlying biological state
$\sigma $: the two cases, where all observations agree
$S=(1,\ldots,1)$ and $S=(0,\ldots,0)$ can be mapped to the biological
states $\sigma_\mathbf{1}$ and $\sigma_\mathbf{0}$ respectively, which
describe ``pure'' states without variation. The remaining
$N - 2$ observable states $S$, where the individual measurements
disagree, correspond to biological states $\sigma$ with random
variation. For the application in our biological study with supposedly
{\em identical} biological systems contributing to the observable
states $S$, the exact pattern leading to contradicting observations
does not carry any information, as long as we assume that there are no
sample specific errors.  Therefore, we subsummize all $N-2$ possible
observations as one biological state $\sigma _\mathbf{r}$ with a
random variation.

The biological rationale for this model is given by the following
example: If one considers a biological variety such as cells in the
retina of the eye, then a certain number of crucial genes ought to be
expressed in all samples. Such genes might include rhodopsin, a
molecule that responds to light. In contrast, genes such as the
hemoglobin family, which are typical of erythrocytes, ought not to be
expressed in the retina. A third class of genes could be considered as
independent of the system in the sense that their expression is not
directly related to the biological system. Such genes may vary in
expression both due to environmental and genetic differences between
the samples.

The model discussed so far is depicted graphically in the left part of
Figure~\ref{noiseline}, where a possible distribution of the relative
frequencies of the three biological states is depicted, for the case
of $m=4$ samples. The distribution can be described by three numbers:
the probabilities $P(\sigma_\mathbf{1})$ and $P(\sigma_\mathbf{0})$,
which contribute to the frequencies of the states $S=(1,1,1,1)$ and
$S=(0,0,0,0)$, and $P(\sigma_\mathbf{r})$ which contributes to both
the frequency of mixed states and the two states above. Describing the
mixed states with only one parameter $P(\sigma_\mathbf{r})$ implies
that the biological variation is modeled evenly and identically
distributed independently for each sample.  In a second step, the
measurement process with possible sample specific errors is modeled as
statistically independent between samples. For each sample $i$, we
define two parameters, $P^i_{0\rightarrow 1}$ and $P^i_{1\rightarrow
  0}$, denoted sample specific {\em error probabilities}. 

To introduce the full formalism of our current model we start by
considering a simple example, again for $m=4$ samples. An
observed state $S$, $S \equiv (S_1,S_2,S_3,S_4)=(1,0,1,0)$, may be
generated by the gene being in state $\sigma_\mathbf{1}$ with the
probability:
\begin{displaymath}
P(\sigma_\mathbf{1})
(1 - P^1_{1 \rightarrow 0})
(P^2_{1 \rightarrow 0}) 
(1 - P^3_{1 \rightarrow 0})
(P^4_{1 \rightarrow 0}), 
\end{displaymath}
or it may be generated by the gene being in state $\sigma_\mathbf{0}$
with the probability:
\begin{displaymath}
P(\sigma_\mathbf{0})
(P^1_{0 \rightarrow 1})
(1 - P^2_{0 \rightarrow 1}) 
(P^3_{0 \rightarrow 1})
(1 - P^4_{0 \rightarrow 1}), 
\end{displaymath}
or it may be generated by the gene being in state $\sigma_\mathbf{r}$
with the probability:
\begin{displaymath}
\begin{array}{l}
P(\sigma_\mathbf{r}) \times \\
\frac{1}{2} [ (P^1_{0 \rightarrow 1}) + (1 - P^1_{1 \rightarrow 0}) ]
\frac{1}{2} [ (1 - P^2_{0 \rightarrow 1}) + (P^2_{1 \rightarrow 0}) ] \times \\ 
\frac{1}{2} [ (P^3_{0 \rightarrow 1}) + (1 - P^3_{1 \rightarrow 0}) ]
\frac{1}{2} [ (1 - P^4_{0 \rightarrow 1}) + (P^4_{1 \rightarrow 0}) ] .\\
\end{array}
\end{displaymath}
With the briefer notation, 
\begin{displaymath}
\begin{array}{lcl}
\rho^i_{1 \rightarrow 0} &  \equiv & 
P^i_{1 \rightarrow 0}\delta_{S_i ,0} + 
(1 - P^i_{1 \rightarrow 0})\delta_{S_i ,1} \\
\rho^i_{0 \rightarrow 1} &  \equiv & 
P^i_{0 \rightarrow 1}\delta_{S_i ,1} + 
(1 - P^i_{0 \rightarrow 1})\delta_{S_i ,0} \\
\end{array}
\end{displaymath}
,where $\delta$ refers to the Kronecker delta (i.e. $\delta_{j,k} = 1$
if $j=k$ and $0$ otherwise), we may express the distribution of
observed states, in the general case of binary discretization with $m$
samples, as:
\begin{equation}
\begin{array}{rcl}
P(S) & = & 
P(\sigma_\mathbf{1})\prod_{i=1}^{m}\rho^i_{ 1\rightarrow 0}  + 
P(\sigma_\mathbf{0})\prod_{i=1}^{m}\rho^i_{ 0\rightarrow 1}  + \\
 & & P(\sigma_\mathbf{r})
\prod_{i=1}^{m}\frac{1}{2}(\rho^i_{ 1\rightarrow 0} + \rho^i_{ 0\rightarrow 1})
\end{array}
\label{discmiscmod}
\end{equation}

{\sloppy Altogether the model uses $3 + 2 * m$ variables.  These
parameters $P(\sigma_\mathbf{1})$, $P(\sigma_\mathbf{0})$,
$P(\sigma_\mathbf{r})$ and $\{P^{i}_{1 \rightarrow 0}, P^{i}_{0
\rightarrow 1}\}_{i=1}^{m}$ are estimated from the observed
distribution of states (right side of Figure~\ref{noiseline}) by
Levenberg-Marquardt \cite{marq63} chi-square minimization of the
unweighted error to the theoretical distribution Eq.\
(\ref{discmiscmod}). Using Eq.\ (\ref{maineq}), and the parameters
estimated as above, our belief that a gene belongs to the underlying
states $\sigma_\mathbf{0}$, $\sigma_\mathbf{1}$, $\sigma_\mathbf{r}$,
given the $2^4=16$ observable states $S$, can now be expressed as: \\
\[  \begin{array}{l}
P(\sigma_\mathbf{0}|S)=\frac{P(S|\sigma_\mathbf{0})P(\sigma_\mathbf{0})}{P(S)}\\
P(\sigma_\mathbf{1}|S)=\frac{P(S|\sigma_\mathbf{1})P(\sigma_\mathbf{1})}{P(S)}\\
P(\sigma_\mathbf{r}|S)=\frac{P(S|\sigma_\mathbf{r})P(\sigma_\mathbf{r})}{P(S)}\\
\end{array}  \]
}

Once the probability  that a gene is in a certain biological state 
$\Sigma^i \in \sigma_\mathbf{1}$, $\sigma_\mathbf{0}$, $\sigma_\mathbf{r}$
 has been calculated for all varieties
$i = 1 \ldots v$, one can calculate the probability
 that a gene exhibits a certain expression
profile over a set of different varieties by taking the product
\beq
 P( \Sigma^1, \ldots, \Sigma^v | S^1, \ldots , S^v)  = 
 \prod_{i=1}^{v} P(\Sigma^i | S^i)
\label{softeq}
\eeq

In this way, the probabilistic state analysis also generates a
clustering: For a given expression profile over the varieties, e.g.
$\sigma_\mathbf{0}^1 \sigma_\mathbf{r}^2 \ldots \sigma_\mathbf{1}^i$,
we may extract those genes for which this expression profile is the
most probable. In fact this is a ``soft'' clustering, in that an
expression profile can belong to several clusters simultaneously with
different probabilities. Moreover the genes clustered to a
biologically interesting expression profile can be ranked by the
probability of Eq.\ (\ref{softeq}).

\subsection{Experimental data preparation}
All cells were grown in RPMI medium supplemented with 7.5\% fetal calf
serum, 10 mM HEPES, 2 mM pyruvate, 50 mM 2-mercaptoethanol and 50 mg
gentamicin per ml (complete RPMI media) (all purchased from Life
Technologies AB, T=E4by, Sweden) at 37=B0C and 5\% CO2.
RNA was prepared using Trizol (GIBCO) and 7.5 =B5g of total RNA was annealed
to a T7-oligo T primer by denaturation at 70=B0C for 10 minutes followed by
10 minutes of incubation of the samples on ice. First strand synthesis was
performed for 2 hours at 42=B0C using 20 U of Superscript Reverse
Transcriptase (GIBCO) in buffers and nucleotide mixes according to the
manufacturers instructions. This was followed by a second strand synthesis
for 2 hours at 16=B0C, using RNAseH, E coli DNA polymerase I and E coli DNA
ligase (all from GIBCO), according to the manufacturers instructions. The
obtained double stranded cDNA was then blunted by the addition of 20 U of
T4 DNA polymerase and incubation for 5 minutes at 16=B0C. The material was
then purified by Phenol:Cloroform:Isoamyl alcohol extraction followed by
precipitation with NH4Ac and Ethanol. The cDNA was then used in an in vitro
transcription reaction for 6 h at 37 =B0C using a T7 IVT kit and biotin
labeled ribonucloetides. The obtained cRNA was purified from
unincorporated nucleotides on a RNAeasy column (Qiagen). The eluted cRNA
was then fragmented by incubation of the products for two hours in
fragmentation buffer (40 mM Tris-acetate, pH 8.1, 100 mM KOAc, 150 mM
MgOAc). 20 =B5g of the final fragmented cRNA was then hybridized to
affymetrix chip U74Av2 (Affymetrix) in 200 =B5l hybridization buffer (100 mM
MES-buffer, pH 6.6, 1 M NaCl, 20 mM EDTA, 0.01
Herring sperm DNA (100 =B5g/ml) and Acetylated BSA (500 =B5g/ml) in an
Affymetrix Gene Chip Hybridization oven 320. The chip was then developed by
the addition of FITC-streptavidin followed by washing using an Affymetrix
Gene Chip Fluidics Station 400. Scanning was performed using a Hewlett
Packard Gene Array Scanner.

 \section{Results}
To evaluate the method we used both real and synthetic data. The
experimental data was generated with Affymetrix microarrays for the
study of differentiating murine B-cells at different stages in the
differentiation process. In this publication the data is only used to
demonstrate the feasibility of the proposed method. The biological
implications of this study are published elsewhere \cite{bcell}.

\subsection{Synthetic data and the effect of correlation}

\showhere{\figcorrbio}

\showhere{\figcorrtran}

For synthetic data, generated with the model parameters\footnote{
These values were chosen as typical values from the estimates on real
data. See next section}, $P(\sigma_\mathbf{0})=0.45$,
$P(\sigma_\mathbf{1})=0.35$, $P(\sigma_\mathbf{r})=0.2$ and
$P^i_{1\rightarrow 0} = P^i_{0\rightarrow 1} = 0.02$ for all samples
$i$, parameter estimates are, as expected, given with low errors. This
result was verified for sample sizes $m=4$, $m=5$, and $m=6$ (data not
shown).

An assumption of simple model used to derive Eq.\ (\ref{discmiscmod})
is that randomly varying genes vary {\em independently} in the samples
of a variety. Hence we investigated how severely this assumption
influences the estimation of the model parameters.
 
To assess the influence of correlations between randomly varying genes
we generated a data set consisting of four bits, i.e. samples, with
the same parameters as above.  In the random patterns a correlation
was introduced between the third and fourth bit by changing the value
of of the fourth to that of the third with a certain probability. We
define this probability as the correlation factor. The correlation was
introduced before distorting the patterns with error probabilities. We
then plotted the mean error in the estimation of parameters over 500
runs of synthetically generated data for correlation factors in the
range $\{0,0.02,\ldots,0.98\}$.

Figure~\ref{globals} shows the error in the estimation of the
parameters describing the underlying distribution. We notice that
even for fully correlated patterns the estimation error is less than
20\% of the correct values.  The estimation of the probability for
biologically varying genes is somewhat worse, for fully correlated
patterns the error is almost 50\%. For real data one can, however,
expect a much smaller correlation.  The average error in the estimates
of the error probabilities, as seen in Figure~\ref{flips}, shows the
expected behavior: The average error grows with the correlation for
the uncorrelated samples, while the estimate for the correlated
observations is almost unaffected. Intuitively, the model compensates
for the correlation by increasing $P(\sigma_\mathbf{1})$ and
$P(\sigma_\mathbf{0})$ as well as the error probabilities and lowering
$P(\sigma_\mathbf{r})$. For correlation factors above $0.50$, due to
the compensation effect, the model deteriorates in explaining the
data. This can be seen in the sum
$P(\sigma_\mathbf{0})+P(\sigma_\mathbf{r})+P(\sigma_\mathbf{1})$ which
initially drops from almost $1$ to $0.99$ as the correlation factor
rises from $0$ to $0.50$ and then from $0.99$ to $0.96$ for
correlation factors in the range $0.50$ to $0.98$ (data not shown).
We hence conclude that it is reasonable not to impose the condition
$P(\sigma_\mathbf{0})+P(\sigma_\mathbf{1})+P(\sigma_\mathbf{r})=1$ in
the model, as this sum indicates if samples are strongly correlated
in genes whose expression vary around the threshold of discretization.

In summary, for not too large correlations in the biological variance
the  algorithm  gives a good quantitative estimate of the model
parameters.  In the case of large correlations the qualitative picture 
given by the estimated parameters is still reliable. 

\subsection{Real data}
Differentiating B-cells are characterized by phenotypic markers into
different stages of development. Here we chose to study the
expressional differences between three such stages; pro, pre and
mature B-cells. For each of these three varieties we used four
different cell lines arrested at the corresponding stage of
development. Measurements we performed with Affymetrix array
containing probesets for 12488 genes and ESTs on each sample. The
discretization of expression levels was given by the Affymetrix
GeneChip absent present calls \cite{affysoft}.

Our algorithm was used to estimate the parameters
$P(\sigma_\mathbf{1})$, $P(\sigma_\mathbf{0})$ and
$P(\sigma_\mathbf{r})$, describing the biological distribution and the
error probabilities (see Table~\ref{params}).  Theoretically, one
expects the three biological probabilities to sum up to one. In our
model, Eq.\ (\ref{discmiscmod}), we do not explicitly impose this
condition. Nevertheless, the sum of the independently estimated
parameters is close to one. This indicates that our model is a
reasonable approximation of the biological system and the measurement
process.

The error probabilities from Eq.\ (\ref{discmiscmod}) can be used as a
consistency index for the samples in a given variety. In the last
variety (mature B-cells) the maximum error probability is notably
higher. This effect is likely to be explained by the different
anatomical origins of the cell lines representing this group. No such
differences exist in the other groups since they all originate in the
bone marrow which is the only anatomical site for B cell development
in the adult animal \cite{gbrm98}.  In contrast, the mature B cell can
reside in several other sites such as spleen, lymph-nodes and
intestine which may affect the gene expression profile in these cells
\cite{rma99, rbycam99}. With only four samples, it is not unlikely
that these effects show up in the error probabilities and not only in
the random variation parameter $P(\sigma_\mathbf{r})$.

\showhere{\tabletyp}

\subsection{Comparison to conventional \\ $t$-test on known genes}
\showhere{\tablettest}

To determine how well biologically relevant information can be
extracted from the discretized data, we compare it with another
statistical method based on continuous expression values.  We use our
method to identify differences in gene expression between two
varieties in the following way. A gene that goes up between variety
$i$ and variety $j$ is characterized by the states
$\sigma^i_\mathbf{0}$, $\sigma^j_\mathbf{1}$ or $\sigma^i_\mathbf{0}$,
$\sigma^j_\mathbf{r}$ or $\sigma^i_\mathbf{r}$, $\sigma^j_\mathbf{1}$.
Hence the belief that a gene goes up is given by the
probability\footnote{Suppressing the conditional probabilities,
$P(\cdot|S)$, for brevity}:
\begin{displaymath}
\begin{array}{l}
P(\mbox{up between } \mbox{variety  $i$ and $j$}) = \\
 \quad P(\sigma^i_\mathbf{0}) P(\sigma^j_\mathbf{1})
+ P(\sigma^i_\mathbf{0}) P(\sigma^j_\mathbf{r})
+ P(\sigma^i_\mathbf{r}) P(\sigma^j_\mathbf{1})
\end{array}
\end{displaymath}
Similarly, the belief that a gene goes down is given by the
probability:
\begin{displaymath}
\begin{array}{l}
P({\mbox{down between variety  $i$ and $j$}}) = \\
\quad P(\sigma^i_\mathbf{1}) P(\sigma^j_\mathbf{0})
+ P(\sigma^i_\mathbf{1}) P(\sigma^j_\mathbf{r})
+ P(\sigma^i_\mathbf{r}) P(\sigma^j_\mathbf{0})
\end{array}
\end{displaymath}
Taking $1-P(\mbox{up})$ thus yields the Bayesian $p$-value of a gene
going up. To answer the same question when working on continuous
expression data one possibility is to employ a one sided two sample
$t$-test in the Welch approximation of unknown variances in the
varieties. This enables testing, for each gene, whether the mean of
expression is higher or lower in one variety than in another. For
comparison of these two approaches we selected a set of genes based on
their well documented expression pattern and biological functions in
the developing B lymphocyte \cite{gbrm98, ls99}. Several of these are
functionally linked since they participate directly in somatic DNA
rearrangement events occurring specifically at the pre-B cell stage or
participate in the regulation of genes involved in this process and
thus display restricted expression patterns (pre-B specific). A second
set of genes were selected based on their expression in cells that are
either committed to the B lineage (B-lineage specific genes, in pre-B
and B-cells) or non committed to this developmental pathway (Not in
B-lineage, expressed in pro-B cells) \cite{rsbnm00}.

The result of this comparison is presented in Table~\ref{Hugo}. For 14
out of the 22 genes the two methods completely agree. Out of these 14
only one (Mb1) does not match the expected target profile. For the
other 8 genes, where the two methods yield different results, the
probabilistic state analysis gives the expected answer in $5$ cases,
which should be compared to the two cases, where the $t$-test gives
the right answer. In one case (rag-1), neither of the two methods gives
the expected result.

For the subset of genes considered here, our method has an advantage
of $5 : 2$ in giving the correct (i.e. expected) expression
pattern. However, the number of samples is not big enough to draw firm
conclusions from this result.

\section{Conclusions} 
The method we have presented serves several purposes:

\begin{enumerate}
\item{It gives a measure of the biological variation of the
genes' expression in different varieties.}
\item{It estimates each hybridizations' global error
probabilities.  These parameters are very useful as they serve as
quality/consistency indicators of the samples of each variety.}
\item{Given the parameters above, it estimates the probability of a
gene belonging to each of the three groups~
$\sigma_\mathbf{0}$, $\sigma_\mathbf{r}$ and $\sigma_\mathbf{1}$. These
probabilities in turn indicate weather the gene is likely to be below,
fluctuating around or above the threshold of discretization.}
\item{Clustering of genes to expression profiles over a set of
different varieties is achieved with Eq.\ (\ref{softeq}). The
probability, i.e. belief, that a gene belongs to a certain cluster is
exactly quantified.}
\end{enumerate}

This novel approach is proven valuable for quantifying both data
reliability and underlying gene expression in microarray
experiments. Our method has been successfully applied in two different
projects \cite{imtech} \cite{bcell}.

\section{Authors' Contributions}
SB and TB have contributed to the development of the model and
implemented the algorithms. MS has contributed the experimental data
and biological expertise.

\vspace{4mm}

{\bf Acknowledgments:} SB and TB are supported by Knut and Alice
Wallenberg Foundation through the SWEGENE consortium. MS is supported
by the swedish science council and the A. \"Osterlund foundation. The
authors also extend their gratitude to Carsten Peterson, Patrik Ed\'en,
Jari H\"akkinen and Markus Rign\'er, for fruitful discussions and careful
proofreading of the manuscript.


\onecolumn

\section{Figures}

\fignoiseline \vfill

\forget{
\figinfoloss \vfill
}

\figcorrbio \vfill

\figcorrtran \vfill
 
\newpage

\section{Tables}

\subsection{Typical paramter values}

\tabletyp \vfill 

\newpage

\subsection{$t$-test vs. probabilistic analysis of gene expression levels}

\tablettest

\end{document}